\documentclass[conference]{IEEEtran}
\IEEEoverridecommandlockouts

\usepackage{fancyhdr}
\renewcommand{\thepage}{\Roman{page}} 

\usepackage{cite}
\usepackage{amsmath,amssymb,amsfonts}
\usepackage{algorithmic}
\usepackage{graphicx}
\graphicspath{{images/}}
\usepackage{textcomp}
\usepackage{xcolor}
\usepackage{booktabs}
\usepackage{multirow}

\newcommand{\R}{\mathbb{R}}

\begin{document}

\title{Quantitative Analysis of Media Bias and Stock Price Dynamics: The 2020 Shock}

\author{\small
\begin{minipage}[t]{0.48\textwidth}
\centering
Shivansh Verma \\
\textit{Dept of Computer Science} \\
\textit{Ashoka University} \\
Sonipat, India \\
shivansh.verma\_ug2023@ashoka.edu.in
\end{minipage}
\hfill
\begin{minipage}[t]{0.48\textwidth}
\centering
Soham Tulsyan \\
\textit{Dept of Computer Science} \\
\textit{Ashoka University} \\
Sonipat, India \\
soham.tulsyan\_ug2023@ashoka.edu.in
\end{minipage}
\\[0.6em]
\begin{minipage}[t]{0.48\textwidth}
\centering
Sashwat Dhanuka \\
\textit{Dept of Computer Science} \\
\textit{Ashoka University} \\
Sonipat, India \\
sashwat.dhanuka\_ug2023@ashoka.edu.in
\end{minipage}
\hfill
\begin{minipage}[t]{0.48\textwidth}
\centering
Anirban Sen \\
\textit{Dept of Computer Science} \\
\textit{Ashoka University} \\
Sonipat, India \\
anirban.sen@ashoka.edu.in
\end{minipage}
}

\maketitle

\thispagestyle{empty}

\begin{abstract}
Whether financial news influences stock prices or simply reflects information already
incorporated into them remains an open question in financial economics. The COVID-19
pandemic provides an opportunity to revisit this question, as it disrupted both news
coverage and financial markets on an unprecedented scale. Existing studies have largely
approached the problem through aggregate sentiment measures, leaving it unclear whether
the observed relationships also hold at the level of individual firms.

We study this question using $6.28$ million news headlines covering $26$ large United
States firms between 2015 and 2025. After filtering the corpus to retain materially
relevant firm-specific coverage, we construct daily stance measures and examine how
their relationship with stock returns changed around the 2020 shock using panel
regressions and vector autoregressions with data-driven structural breaks.

Our findings indicate that the relationship between financial news and equity markets is
more nuanced than aggregate analyses alone suggest. While we find little evidence of a
persistent market-wide change in media stance or stock returns following the pandemic,
dynamic relationships emerge for a subset of firms around their own structural breaks.
Taken together, these results suggest that understanding media--market interactions
requires firm-specific analysis and provide a framework for studying how news and prices
interact under changing market conditions.
\end{abstract}

\section{Introduction}
The news is one of the primary channels through which information about a company
reaches the investors who buy and sell its stock, and for that reason the tone of a
firm's coverage has long been suspected of moving its share price. When a national
outlet casts a firm's quarter as a triumph or its leadership as contended, the framing
travels to thousands of readers who already hold the stock or might soon buy it, and
whether that framing leaves a mark on the price or merely repeats what the market has
already worked out is a question that reaches well beyond the newsroom. It matters to
the investor deciding whether a run of hostile headlines is a signal worth trading on,
to the firm weighing how much its public image is worth defending, and to the regulator
asking whether the press can tilt the fortunes of the largest companies. The question
is also stubbornly hard to answer, because coverage and prices may move together for many
reasons that have nothing to do with one driving the other, and any convincing account
has to pull apart the coverage that carries genuine new information from the coverage
that only narrates a move the market has already made.

The sharpest way to draw that line is the efficient market hypothesis~\cite{fama1970},
which holds that the price of a stock at any given moment already reflects every piece
of information publicly available about the firm. Read strictly, this idea leaves no
room for the tone of a headline to forecast returns, because whatever the headline
reports has been folded into the price by the time it appears in print, and coverage
becomes a mirror held up to the market rather than a force acting on it. The hypothesis
matters to us not because we expect it to hold to the letter but because it sets the
null against which any claim of media influence has to be measured. If the tone of a
firm's coverage does help predict its returns, then either the market is slow to absorb
public information or the coverage is carrying something the market has not yet seen,
and if the tone predicts nothing, then the press is largely reporting the market back to
itself. We take no position in advance on which of these holds, and we do not even
assume that influence, where it exists, runs from coverage to prices rather than the
other way around, since a firm whose stock has just tumbled tends to draw darker
coverage in the days that follow.

The year 2020 offers a rare occasion to put these questions to a genuine test. The
onset of the pandemic was a shock that fell on every firm at once, crowded out almost
everything else in the news for months, and repriced the whole market within a few
violent weeks, so if the relationship between coverage and returns were ever going to
bend out of its usual shape, this is the moment it would show. We treat the shock as a
natural experiment and ask three questions of it. The first is whether the tone of a
firm's coverage stepped to a new level once the shock arrived. The second asks the same
of prices, whether firm returns shifted to a new level after the shock. The third asks whether the relationship between tone and returns changed its
behavior around 2020 even when neither series shifted its average, and in particular
whether coverage began to lead returns, or returns to lead coverage, in a way it had
not before.

Answering these questions calls for a measure of tone that belongs to a single firm
rather than to the market as a whole, so we begin by reading each headline with a
target-dependent sentiment model that scores how favorably the headline speaks about the
specific firm it names~\cite{hamborg2021}, and we collapse those readings into one
signed daily figure for each company. For the first two questions we place that figure
and the firm's returns on a panel regression with fixed effects for every firm and
every day. This is because a plain before-and-after comparison of these distributions would confuse a genuine 2020 break with two things that have nothing to do with the
pandemic. This could be due to factors like some firms being more heavily and more
favorably covered than others, and market rallies that push every firm in the same
direction on a given day. Fixed effects are used here as they hold both of these aside and leave the
regression looking only at how a firm changed relative to its own past once the day's
common movement has been removed. For the third question a single regression is not
enough, because the direction of any link is precisely what we want to learn rather than
assume, so we turn to a vector autoregression. This treats tone and returns as jointly determined, through which the model allows the data to determine whether predictive relationships run from tone to returns, from returns to tone, or in neither direction.

The results are less dramatic than the 2020 shock might suggest, but that lack of a strong effect is the main finding.
 From an initial pull of $6.28$ million headlines we
distilled a corpus of $90{,}579$ materially relevant, firm-scored headlines across $26$
large United States firms. From that corpus neither the tone of coverage nor firm
returns shows a detectable shift in level after the shock. The relationship between the two turns out to be real but local.
Where coverage and returns do lead one another, the link belongs to individual firms. This effect tends to show only after each firm's own structural break, and once the firms are
pooled, no market-wide channel survives in either
direction. We find little evidence that the 2020 shock fundamentally changed the relationship between the financial press and the market. Instead, the relationship remained heterogeneous across firms before and after the crisis. Because our approach accounts for common shocks, it avoids overstating a market-wide effect suggested by the timing of the crisis.

\section{Related Work}

\subsection{Media Sentiment and Market Outcomes}
A large recent literature reports that the tone of financial text moves with market
outcomes, and most of it measures that tone for a market rather than for a firm.
Costola et al.~\cite{costola2023} score $203{,}886$ pandemic news articles from three
outlets with a finance-adapted transformer and find that a more positive tone is
associated with higher index returns. Bai et al.~\cite{bai2023} reach a similar
conclusion on $1.29$ million texts across $47$ countries, and add that negative
sentiment moves returns more than positive sentiment does. Huynh et
al.~\cite{huynh2021} build a feverish-sentiment index for $17$ economies and show that
it predicts volatility positively and returns negatively at the onset of the crisis.
A parallel group constructs standing indices from search behaviour rather than text.
Anastasiou et al.~\cite{anastasiou2022} form a positive COVID-19 index from Google
search volume and find that it cushions return declines across G20 markets, and later
tie vaccine-related search to falls in policy uncertainty and market
fear~\cite{anastasiou2026}. Others weigh one channel against another. Verma and
Verma~\cite{verma2025} find that economic-news sentiment carries more return-relevant
information than the noisier social-media stream, and Eierle et al.~\cite{eierle2022}
show that a social-media proxy predicts short-horizon returns once fundamentals are
held fixed.

Some of this work already suggests that the effect is uneven across firms. Kim-Hahm et
al.~\cite{kimhahm2025} track three large technology firms at weekly frequency and find
that sentiment moves volume and volatility differently for each. Sing and
Singh~\cite{sing2023} report sector- and wave-dependent responses in India, and Dong et
al.~\cite{dong2022} find that media sentiment moves retail-driven first-day returns in
a high-growth market and moves them more during the pandemic than before it. Chen et
al.~\cite{chen2022} show that firm-specific return variation rises with a country's
level of digital development, which implies that firm-level information is exactly
what an aggregate index discards. Two studies come closest to our own measurement.
Mahmoudi et al.~\cite{mahmoudi2022} tie firm-specific sentiment to corporate
announcement returns, and Liu et al.~\cite{liu2025} combine firm-level and
industry-level sentiment over $1.39$ million texts to improve credit-spread forecasts,
though their outcome is a bond spread rather than an equity return. Where tone is used
predictively at the firm level, the design is usually a short-horizon classifier rather
than an econometric test~\cite{ruan2025}. Across the strand, the question of whether
coverage of a firm predicts that firm's return, in a form that can be tested against a
null, is not posed.

\subsection{Measuring Tone at the Level of a Firm}
Assigning tone to the correct firm is a separate research problem, and the tools have
been moving toward it. Domain adaptation came first, since models pretrained on
financial text read sentiment more accurately than general-purpose
ones~\cite{araci2019, liu2020, priya2025}. Attribution to a named entity came next.
Tang et al.~\cite{tang2023} annotate entity spans in financial news and find that
supervised entity-level models outperform zero-shot prompting. Rønningstad et
al.~\cite{ronningstad2022} survey the entity-level task and show that document-level
classifiers resolve sentiment incorrectly when a text names several entities, and
Daudert~\cite{daudert2022} builds a multi-source corpus annotated for both sentiment
and relevance for the same reason. Gyawali et al.~\cite{gyawali2025} extend targeted
stance detection to filings and earnings calls. Van der Heever et al.~\cite{heever2026}
go further and subject aspect-level sentiment to placebo and stability tests, finding
that many reported sentiment-return associations do not survive them.

The same literature reports that the measurement is fragile. General-purpose models
underperform fine-tuned domain models on financial tasks~\cite{li2023}, and different
architectures suit different tasks~\cite{kirtac2025}. Reasoning prompts make sentiment
classification worse rather than better~\cite{vamvourellis2025}, benchmarks on hedged
financial language expose further reliability limits~\cite{kubica2025}, the sentiment
of the prompt itself shifts model output~\cite{gandhi2025}, and models adopt the
framing of the analyst reports they are shown~\cite{hu2026}. Point-in-time testing
shows that apparent predictive skill can be memorisation of the sample
period~\cite{benhenda2026, eliseev2026}. We treat these results as design constraints.
Every score in this paper comes from one target-dependent stance
model~\cite{hamborg2021} placed behind one fixed relevance stage, so that no later
result depends on a model choice, a prompt, or information that was unavailable when
the headline was published.

\subsection{Dynamic Links and Structural Change}
A third strand asks which direction the relationship runs, and whether it held through
2020. Conforti et al.~\cite{conforti2022} show the link is strong enough to run
backwards, using intraday prices to improve stance detection on merger tweets, which is
our question with the direction reversed. Where causality is tested forward, it is
usually tested on aggregate series. Moutinho et al.~\cite{moutinho2025} find
bidirectional, time-varying causality between an uncertainty-based sentiment measure
and S\&P~500 spreads that is strongest during the pandemic, and Cevik et
al.~\cite{cevik2022} reach comparable conclusions for G20 markets using panel fixed
effects and a panel VAR, though their sentiment proxy is search volume rather than
news text. Mamaysky~\cite{mamaysky2024} studies the news-market relationship through
2020 directly and finds it does not hold its earlier form. Smith and
Yamagata~\cite{smith2011} bring the analysis down to the firm using dynamic panels, but
carry no text variable at all.

The econometric literature supplies both the tools and the warning. Investor-sentiment
series contain endogenous structural breaks, and ignoring them distorts any model built
on top~\cite{ballinari2020}; the same is true of the volatility series on the price
side~\cite{oliveira2024}. Karavias et al.~\cite{karavias2023} detect and date breaks in
a firm panel and apply the procedure to the stock-market reaction to COVID-19, and
related work extends break detection to interactive effects~\cite{ditzen2025jae,
ditzen2025stata}, to changes in group membership~\cite{lumsdaine2023}, and to settings
where the timing of treatment is itself unknown~\cite{pretis2026}. Rossi and
Wang~\cite{rossi2019} show that Granger tests lose power when parameters drift unless
they are made robust to it. On the price side, the 2020 shock is well documented as a
cross-sectional event: firms differing in resilience to social distancing earned
sharply different returns~\cite{pagano2023}. None of this work asks whether the shock
also changed the relationship between what the press writes about a firm and what its
stock does.

The three strands leave one gap in common. Tone is measured for a market rather than
for a firm, direction is tested on aggregate series, and the break date is taken from
the calendar rather than estimated from the data. A relationship that belongs to
individual firms, and that begins at each firm's own break, cannot be detected under
these choices. We address this by scoring each headline for the firm it names, testing
predictability in both directions for single firms and for the pooled panel, and dating
every break from the series itself.

\section{Data Collection}

\subsection{Corpus and Coverage}
Our object of study is the tone of the news that surrounds large United States
firms, so the corpus has to begin as close as possible to the full stream of that
coverage. We assemble it from MediaCloud, a service that continuously indexes the
output of national news outlets, and we query its United States national collection
one firm at a time over the period 2015 to 2025. Each query returns the news items
that name the firm, each carrying a publication date, a source outlet, and a
headline. We keep the headline as the unit of analysis, because it is the
one part of a news item written to compress the story's angle into a single line,
and it is therefore where a firm's coverage shows its stance most plainly. Seeded with roughly $250$ large-capitalization
United States firms, the raw pull returns $6{,}284{,}404$ headlines.

 We also tested whether GDELT could expand our coverage, but it did not add useful data.
 The headlines GDELT returned were of visibly
poorer quality, often truncated or malformed and heavily duplicated across
near-identical records, so folding them into the corpus would have thinned a clean
signal with noise rather than deepening it. We keep the study on the MediaCloud
stream, whose headlines are consistently well formed and attributed, and treat that
stream as the raw material for everything that follows.

One feature of that raw material shapes every later choice. Coverage is spread
across firms with extreme inequality: a small number of the most heavily reported
names account for the bulk of all headlines, while a long tail of firms surfaces in
the news only occasionally. 
\subsection{Data Preprocessing}

\begin{figure}[t]
\centering
\includegraphics[width=\linewidth]{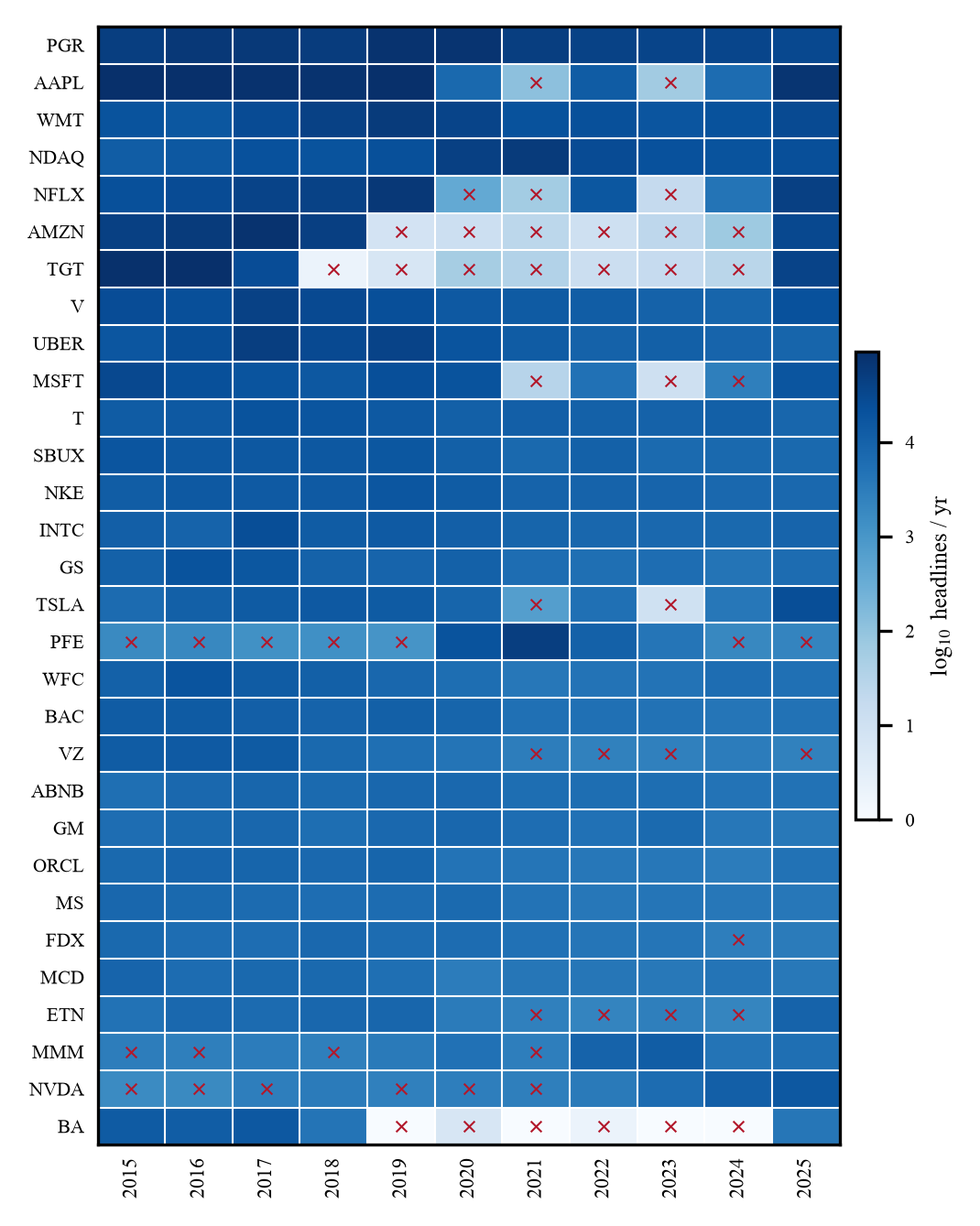}
\caption{Yearly headline coverage for each candidate firm, shaded by
$\log_{10}$ headlines per year. A cross marks any year in which a firm sits at or
below the three-thousand-headline floor; a firm crossed in more than three of the
eleven years is set aside as too intermittently covered to support a continuous
series.}
\label{fig:strat}
\end{figure}


The skew actually provides us with meaningful information; it decides which firms can be
studied at all. A daily, firm-level measure of tone is only as steady as the
coverage beneath it, and a firm the news names a handful of times a month cannot
carry a daily series that means anything. We therefore keep the firms the press
reports on heavily and let the thin ones go. Ranking the universe by the volume of
coverage each firm attracts and holding onto the most reported names concentrates
the corpus onto its densely covered head, some $4{,}706{,}589$ headlines, and leaves
behind a long tail of firms that never generated enough news to measure.

Volume in a single year is not enough on its own, because coverage also has to
persist across the whole decade: a firm that falls quiet for years leaves stretches
of time that no series can span. We therefore hold each remaining firm to a floor of
more than three thousand headlines in a year, and set aside any firm that slips to or
below that floor in more than three of the eleven years from 2015 to 2025, since a
firm that runs thin often cannot anchor a continuous weekly or daily series.
Figure~\ref{fig:strat} maps yearly coverage firm by firm and marks the years that
fall below the floor; the firms carried forward are those whose rows stay dark almost
throughout, while the ones that appear pale year after year are the ones the rule
removes.

\begin{table}[!htb]
\caption{From the full news scrape to the corpus the study runs on.}
\label{tab:funnel}
\centering
\small
\begin{tabular}{lrr}
\toprule
Stage & Headlines & Firms \\
\midrule
Full MediaCloud scrape & $6{,}284{,}404$ & $250$ \\
Most-covered firms retained & $4{,}706{,}589$ & $30$ \\
Name or ticker in title, de-duplicated & $695{,}731$ & $26$ \\
\bottomrule
\end{tabular}
\end{table}

Two kinds of contamination still exist. A great many headlines
name a firm only in passing, in a list or an aside, while the story itself is about
something else; because a headline written about the firm almost always carries the
firm's name or its ticker symbol, we keep only the headlines that do and discard the
rest. The same wire report, in turn, is republished word for word across dozens of
outlets, so we collapse these near-identical copies to a single record per firm.
Table~\ref{tab:funnel} traces the corpus through these passes, from the full scrape
down to what remains: $695{,}731$ scored headlines across $26$ firms, each attached
to a firm the news followed closely and each naming that firm in its title. 

\section{Methodology}

\subsection{Relevance Classification}
\label{sec:rc}

The scored corpus still carries a problem that name matching alone cannot resolve. A
headline that names a firm is not necessarily about the firm. ``Apple'' headlines a
recipe as easily as an earnings call, banks lend their names to stadiums, and a firm
can appear in a story as a passing reference rather than its principal subject. If such
headlines are included in a firm's daily stance measure, they introduce observations
that do not reflect information about the firm's own activities. We therefore restrict
the corpus to headlines in which the firm is the primary subject of a material corporate
event, such as earnings, mergers and acquisitions, executive changes, large layoffs,
regulatory or legal action, major product launches, or credit and bankruptcy events.
``Goldman Sachs beats profit estimates on a trading surge'' belongs in this set.
``Ten apple recipes to try this autumn'' does not, even though both contain the firm's
name.

\begin{figure}[!t]
\centering
\includegraphics[width=\linewidth]{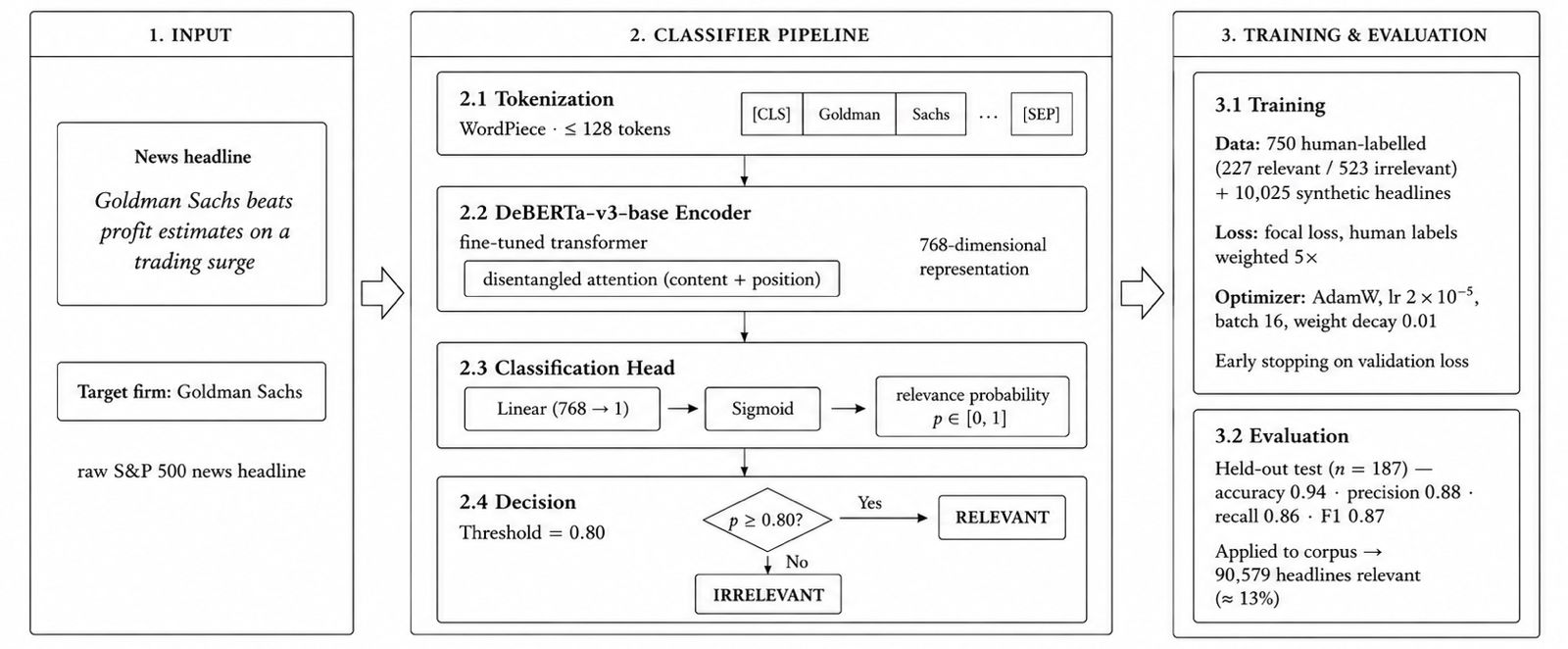}
\caption{The relevance-classifier pipeline. A raw S\&P~500 news headline is tokenized
(at most $128$ tokens) and read by a fine-tuned DeBERTa-v3-base encoder; a linear head
maps the $768$-dimensional representation to a relevance probability, thresholded at
$0.80$ for full-corpus inference. The model is trained on $750$ human-labelled
headlines ($227$ relevant, $523$ irrelevant) and $10{,}025$ synthetic ones under focal
loss, with the human labels weighted five times the synthetic, using AdamW (learning
rate $2\times10^{-5}$, batch size $16$, weight decay $0.01$) and early stopping on the
validation loss. On the held-out test set ($n=187$) it attains an accuracy of $0.94$, a
precision of $0.88$, a recall of $0.86$, and an F1 of $0.87$, and it retains $90{,}579$
headlines (about $13\%$) from the scored corpus.}
\label{fig:rc_pipeline}
\end{figure}

Separating the two at the scale of hundreds of thousands of headlines calls for a
supervised classifier, and a classifier calls for labelled data. We hand-labelled
$750$ headlines against the rubric above, resolving borderline cases through review,
yielding $227$ relevant and $523$ irrelevant examples. A further $187$ labelled
headlines ($145$ irrelevant, $42$ relevant) were reserved as a held-out test set and
used neither for training nor for selecting the decision threshold. The classifier
itself is a fine-tuned DeBERTa-v3-base model, a transformer well suited to the
context-dependent language characteristic of short financial headlines. It maps each
headline to a single probability that the headline is materially relevant to the firm
it names. Figure~\ref{fig:rc_pipeline} summarises the complete pipeline, from a raw
headline through tokenization, the encoder, and the decision threshold, together with
the data and training configuration.

Trained on the $750$ manually labelled headlines alone, the model reaches a precision
of only $0.65$ at a recall of $0.71$. This largely reflects the class distribution in
the training data. The labelled set contains roughly one relevant headline for every
two irrelevant ones, whereas the proportion of relevant headlines in the full corpus is
closer to one in eight, encouraging the model to classify headlines as relevant more
often than is appropriate for the complete dataset. Additional manual annotation would
reduce this mismatch, but only at substantial cost. Instead, we augment the training
set with $10{,}025$ synthetic headlines generated from sector-aware templates, whose
events, amounts, and counterparties remain plausible for each firm's industry. These
examples are concentrated in categories the initial model finds most difficult,
including consumer promotions, analyst previews, immaterial lawsuits, and passing
mentions on the irrelevant side, and earnings, layoffs, and unusually phrased
transactions on the relevant side. Samples of the generated headlines are checked
against the annotation rubric to ensure they are consistent with the intended labels and
do not duplicate the held-out test set. Training then uses focal loss, with every
manually labelled example weighted five times more heavily than a synthetic one. The
synthetic data therefore broaden the range of examples encountered during training while
ensuring that human annotations remain the dominant source of supervision.

The final model achieves an accuracy of $0.94$, a precision of $0.88$, a recall of
$0.86$, and an F1 score of $0.87$ on the held-out test set at the default threshold.
For full-corpus inference we instead adopt a decision threshold of $0.80$, where
precision increases to $0.92$ while recall remains $0.81$. This operating point reflects
the role of the classifier within the subsequent analysis. Every headline classified as
relevant contributes directly to a firm's measured daily stance, whereas an irrelevant
headline admitted to the corpus introduces measurement error into that quantity.
Excluding a genuinely relevant headline reduces the information available on a
particular day, but does not alter the interpretation of the headlines that remain.
Since the former error has more serious consequences for the validity of the stance
measure than the latter, we favour a threshold that prioritises precision over recall.
Applied to the scored corpus, the classifier retains $90{,}579$ headlines,
approximately $13\%$ of the total. These retained headlines form the corpus used
throughout the remainder of the analysis.

\begin{figure}[!t]
\centerline{\includegraphics[width=\linewidth]{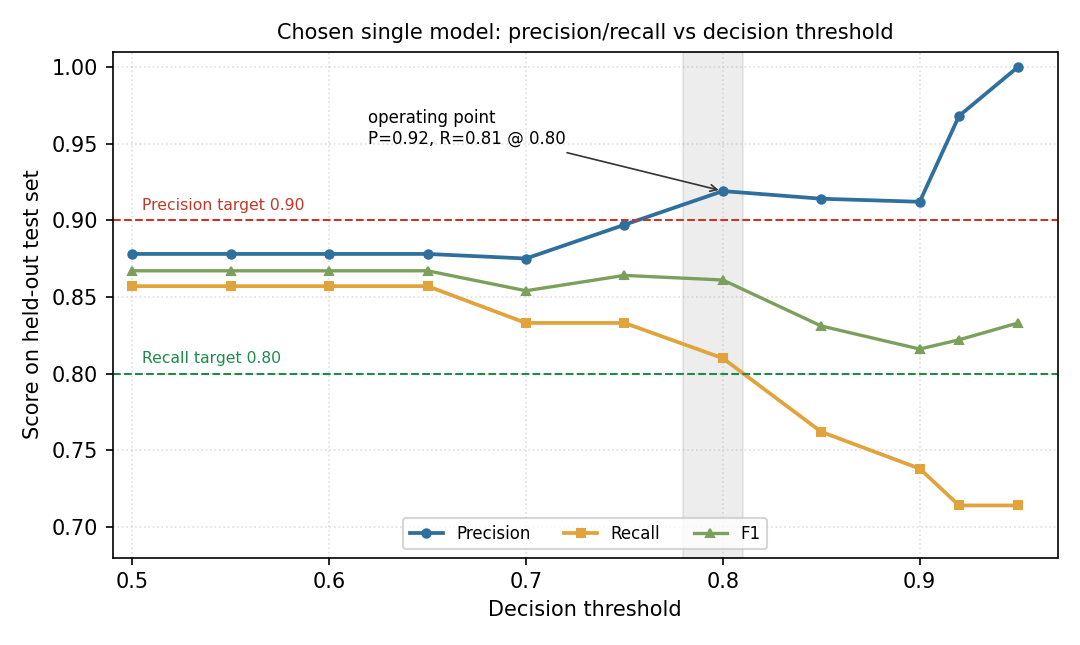}}
\caption{Precision, recall, and F1 of the final relevance classifier on the
held-out test set ($n=187$) as the decision threshold varies. The operating
threshold of $0.80$ trades recall for precision ($P=0.92$, $R=0.81$), the direction
the full-corpus inference favors.}
\label{fig:relsweep}
\end{figure}

\subsection{RQ1: Did Media Stance Shift After the Shock?}
With a daily, firm-level stance series in hand, the first question the 2020 shock
raises is the plainest one: did the tone of a firm's coverage step to a new level once
the shock arrived? The tempting way to answer it is to average each firm's stance over
the years before 2020 and over the years after, then read off the difference. That
comparison would mislead, and seeing why fixes the design for this question and the
one that follows.

Two forces may move a firm's coverage that have nothing to do with the shock, and a raw
before-and-after average folds both of them into what it would label the 2020 effect.
Firms differ from one another in ways that persist across the whole sample, so that
some are covered more, and more warmly, than others. And on any single day every firm
is pushed in the same direction by the news cycle as a whole, so a run of shared bad
days, or a change in which firms happen to be in the news, would surface as a break
that owes nothing to the pandemic. A panel regression with two sets of fixed effects
removes exactly these two confounders. Firm effects tie each firm to its own baseline,
so the comparison is between a firm and its own past rather than between one firm and
another; daily effects hold constant whatever is common to every firm on a given day,
so a market-wide swing in tone cannot be read as a firm-level break. What remains for
a post-2020 indicator to explain is the within-firm change, net of these common shocks,
which is precisely the quantity the question asks about. The estimate is credible
under three assumptions the setting makes reasonable: that in absence of the shock, the pre-
and post-periods would have followed a common trend; that the timing of the break is
set by the pandemic's onset rather than by anything in a firm's own coverage; and that,
once the firm and day effects are removed, what is left is uncorrelated with the post
indicator. We guard the last of these by reporting HC1-robust standard errors and,
because several firms are often covered on the same day, standard errors clustered by
day.

The outcome that carries this question is the tone of a firm's coverage on a day. A
target-dependent sentiment model reads each retained headline and reports, for the
firm the headline names, how likely its tone toward that firm is positive, negative,
or neutral; we write these three probabilities as $p^{+}_a$, $p^{-}_a$, and $p^{0}_a$
for headline $a$, with $p^{+}_a + p^{-}_a + p^{0}_a = 1$. We collapse them into a
single signed score, $p^{+}_a - p^{-}_a$, which runs from $-1$ for wholly unfavorable
coverage to $+1$ for wholly favorable coverage and sits near zero when a headline is
even-handed or the model is unsure. A firm's stance on a day is the mean of these
scores over the headlines it drew that day,
\begin{equation}
\text{bias}_{it} = \frac{1}{N_{it}} \sum_{a \in A_{it}} \bigl(p^{+}_a - p^{-}_a\bigr),
\label{eq:bias}
\end{equation}
where $A_{it}$ is the set of retained headlines about firm $i$ on day $t$,
$N_{it} = |A_{it}|$ is that day's coverage volume, and
$\text{bias}_{it} \in [-1,1]$ is the average tone toward the firm.

To ask whether this daily stance moved to a new level after the shock, we place it on
the post-2020 indicator alongside the firm and daily effects and the day's coverage
volume,
\begin{equation}
\text{bias}_{it} = \alpha_i + \delta_t^{\text{day}} + \beta\,\text{Post}_t
+ \gamma\,\text{vol}_{it} + \varepsilon_{it}.
\label{eq:rq1}
\end{equation}
Here $\text{Post}_t = \mathbf{1}[\,t \ge \text{2020-01-01}\,]$ switches on for every day
from January 2020 onward, and its coefficient $\beta$ is the quantity of interest;
$\alpha_i$ are the firm effects and $\delta_t^{\text{day}}$ the daily effects. The
volume control $\text{vol}_{it} = N_{it}$ enters because a daily average tightens
mechanically as more headlines fold into it, so holding the count fixed keeps a
genuine change in tone from being mistaken for a change in the sheer amount of
coverage. We test
\begin{equation*}
H_0^{(1)}:\ \beta = 0 \qquad\text{against}\qquad H_1^{(1)}:\ \beta \neq 0,
\end{equation*}
estimating \eqref{eq:rq1} by the within-transformation with HC1-robust standard
errors and, since several firms can be covered on the same day, day-clustered standard
errors alongside them.

\subsection{RQ2: Did Firm-Level Returns Shift After the Shock?}
The second question asks the same of prices. The outcome is firm $i$'s daily log
return, $\text{return}_{it} = \ln P_{it} - \ln P_{i,t-1}$, where $P_{it}$ is the
firm's closing price on day $t$; we draw the daily closing prices, the S\&P~500 index,
and the VIX volatility index from Yahoo Finance. Log returns are additive over
time and roughly symmetric, which suits a linear model. Returns carry a large common
component: when the market rises, most firms rise with it, so the design must strip
that component out before it can see a firm-level break. Monthly effects absorb the
broad market regime, and within each month the S\&P~500 return and the VIX account for
the day-to-day market swings and the level of volatility. The regression is
\begin{equation}
\text{return}_{it} = \alpha_i + \delta_t^{\text{month}} + \beta\,\text{Post}_t
+ \gamma_1\,\text{sp500\_ret}_t + \gamma_2\,\text{vix}_t + \varepsilon_{it},
\label{eq:rq2}
\end{equation}
with $\text{Post}_t$ as before, $\alpha_i$ the firm effects, $\delta_t^{\text{month}}$
the monthly effects, and $(\text{sp500\_ret}_t,\,\text{vix}_t)$ the market controls. We
again test the shock,
\begin{equation*}
H_0^{(2)}:\ \beta = 0 \qquad\text{against}\qquad H_1^{(2)}:\ \beta \neq 0,
\end{equation*}
estimating \eqref{eq:rq2} by the within-transformation with HC1-robust standard errors.
Table~\ref{tab:variables} gathers every quantity that enters the two regressions.

\begin{table*}[!t]
\centering
\small
\caption{Variables in the RQ1 (stance) and RQ2 (returns) regressions.}
\label{tab:variables}
\begin{tabular}{p{2.6cm}p{1.4cm}p{8.4cm}}
\toprule
\textbf{Variable} & \textbf{Used in} & \textbf{Meaning} \\
\midrule
$\text{bias}_{it}$ & RQ1 (dep.) & Daily firm stance: the signed mean $p^{+}_a-p^{-}_a$ of the headlines about firm $i$ on day $t$. \\
$\text{return}_{it}$ & RQ2 (dep.) & Daily log return of firm $i$, $\ln P_{it}-\ln P_{i,t-1}$. \\
$\text{Post}_t$ & RQ1, RQ2 & Shock indicator, $\mathbf{1}[t\ge\text{2020-01-01}]$; its coefficient $\beta$ is the parameter of interest. \\
$\text{vol}_{it}$ & RQ1 & Coverage volume: the number of retained headlines about firm $i$ on day $t$. \\
$\text{sp500\_ret}_t$ & RQ2 & Daily S\&P~500 index return, controlling for broad market movement. \\
$\text{vix}_t$ & RQ2 & Daily VIX level, controlling for market volatility. \\
$\alpha_i$ & RQ1, RQ2 & Firm effects, absorbing permanent differences across firms. \\
$\delta_t^{\text{day}}$ & RQ1 & Daily effects, absorbing shocks common to all firms on a day. \\
$\delta_t^{\text{month}}$ & RQ2 & Monthly effects, absorbing shocks common to all firms in a month. \\
\bottomrule
\end{tabular}

\vspace{2pt}
{\footnotesize \textit{Note:} ``(dep.)'' marks the dependent variable; the remaining rows are regressors or fixed effects.}
\end{table*}

\subsection{RQ3: Did the Stance--Return Relationship Change?}
\label{sec:var}
The two level tests leave a question they cannot reach. Learning that neither the
tone of coverage nor firm returns stepped to a new level after the shock says
nothing about whether the two moved together differently once it arrived; a
relationship can change how it behaves while each series holds its average in
place. The third question therefore asks not where stance and returns sit but how
they lead each other, and whether that lead--lag structure shifted around 2020. We
leave the direction of any such link open. Coverage may carry information prices
have yet to absorb, so that past stance forecasts returns, or prices may move first
and coverage follow, so that returns forecast stance.

A vector autoregression answers to exactly that requirement, because it treats
stance and returns as jointly endogenous and reads the direction of predictability
off the data rather than off an assumed ordering. For a firm we gather the two
series into $\mathbf{y}_t = (\Delta\text{bias}_t,\, r_t)^{\top}$ and write the
reduced-form VAR of order $p$ as
\begin{equation}
\mathbf{y}_t = \mathbf{c} + \sum_{\ell=1}^{p}\mathbf{A}_\ell\,\mathbf{y}_{t-\ell}
+ \mathbf{u}_t,
\label{eq:var}
\end{equation}
with coefficient matrices $\mathbf{A}_\ell \in \R^{2\times2}$ and innovations
$\mathbf{u}_t$ of covariance $\Sigma$. The off-diagonal terms hold the cross-dynamics
the question turns on: a nonzero $(2,1)$ entry means past stance moves current
returns, and a nonzero $(1,2)$ entry means past returns move current stance. We fit
the model at a weekly frequency. Daily firm-level stance is too thin for many firms
to support a lag-rich dynamic model, whereas aggregating the headlines into calendar
weeks restores a dense, near-continuous series for each retained firm while leaving
enough weeks for the lags the model needs. We align each weekly stance series to the
firm's weekly log return, bridge the short coverage gaps by imputation, and keep an
observed-only version of every series so that no dynamic result can rest on the
filled values.

Two properties of the weekly data determine the modelling approach. A VAR model can only be reliably applied to stationary time series, i.e, series whose statistical behaviour (such as their average and variance) remains stable over time. If a series is non-stationary (for example, it has a persistent trend), a regression on the raw values can identify relationships that are simply due to the shared trend rather than a genuine connection, resulting in a spurious regression. To ensure the data satisfy the VAR assumptions, we test each series for stationarity using the Augmented Dickey–Fuller (ADF) and KPSS tests. Together, these tests provide strong evidence about whether each time series is suitable for analysis in a levels VAR model. Weekly returns come back stationary for every firm, while the stance
index is stationary for some firms and integrated or break-driven for the rest. To
hold every firm to one specification without running into a spurious regression, we take
the VAR on the first difference of the stance index, $\Delta\text{bias}_t$, which is
stationary throughout, paired with the weekly return; the handful of firms whose
stance and returns test as cointegrated we carry separately as robustness exceptions
to the differenced form.

Because the question is about a structural change, we let the data date the change rather
than impose the calendar. A Bai-Perron procedure locates the single largest shift in
the mean of each weekly series, and that estimated break splits the firm's sample into
a pre- and a post-break segment on which the dynamics can be compared. Dating the
break empirically is what makes the comparison honest: some firms break near the
pandemic window while others break on news of their own, so a single imposed 2020 cut
would misplace the change for many of them. For each firm, we select the VAR lag order by minimizing the Bayesian Information Criterion (BIC) over $p = 1,\dots,8$. The BIC penalizes unnecessary parameters, favouring simpler and more reliable models. We then verify that each VAR is stable and that the residuals exhibit no autocorrelation. All statistical inference is performed using standard errors that are robust to heteroskedasticity and autocorrelation.

On each fitted model we test predictability in both directions through Granger
causality, which asks whether the past of one series helps predict the other beyond
that other's own past. Writing $a^{(2,1)}_\ell$ for the coefficient on lag-$\ell$
stance in the return equation, stance fails to forecast returns when
\begin{equation*}
H_0^{(3)}:\ a^{(2,1)}_1 = \cdots = a^{(2,1)}_p = 0
\end{equation*}
holds against the alternative that at least one of these coefficients is nonzero, and
the mirror-image hypothesis on the $(1,2)$ coefficients asks whether returns forecast
stance. We test each with a joint Wald statistic and Newey--West standard errors that
absorb the residual dependence found above, on the full sample and on the pre- and
post-break segments in turn.

A firm-by-firm test is underpowered, since each rests on one short series, so the
specification the dynamic conclusion leans on pools the firms into a panel VAR.
Pooling buys power but revives a hazard: ordinary within-firm demeaning ties the
transformed errors to the lagged regressors and bends the estimated dynamics, the
Nickell bias. We strip firm heterogeneity with the Arellano--Bover (Helmert) forward
transformation instead, which subtracts from each observation the mean of its own
future values and so keeps the lagged regressors orthogonal to the transformed error,
\begin{equation}
\tilde{y}_{it} = c_{it}\Bigl(y_{it} - \tfrac{1}{T_{it}}\textstyle\sum_{s>t} y_{is}\Bigr),
\end{equation}
with $c_{it}$ a factor that equalizes the variance of the transformed errors. Into the
pooled model we partial the very forces that could otherwise counterfeit a
media--market link: the post-2020 indicator, the VIX, the weekly S\&P~500 return, and
the week's coverage volume. A Granger relation that survives these controls cannot be
an echo of market-wide volatility or of how much was written, which is the standard the
paper's central claim is held to.

\section{Results}

\subsection{RQ1: No Detectable Shift in Media Stance}
\begin{figure}[!t]
\centering
\includegraphics[width=\linewidth]{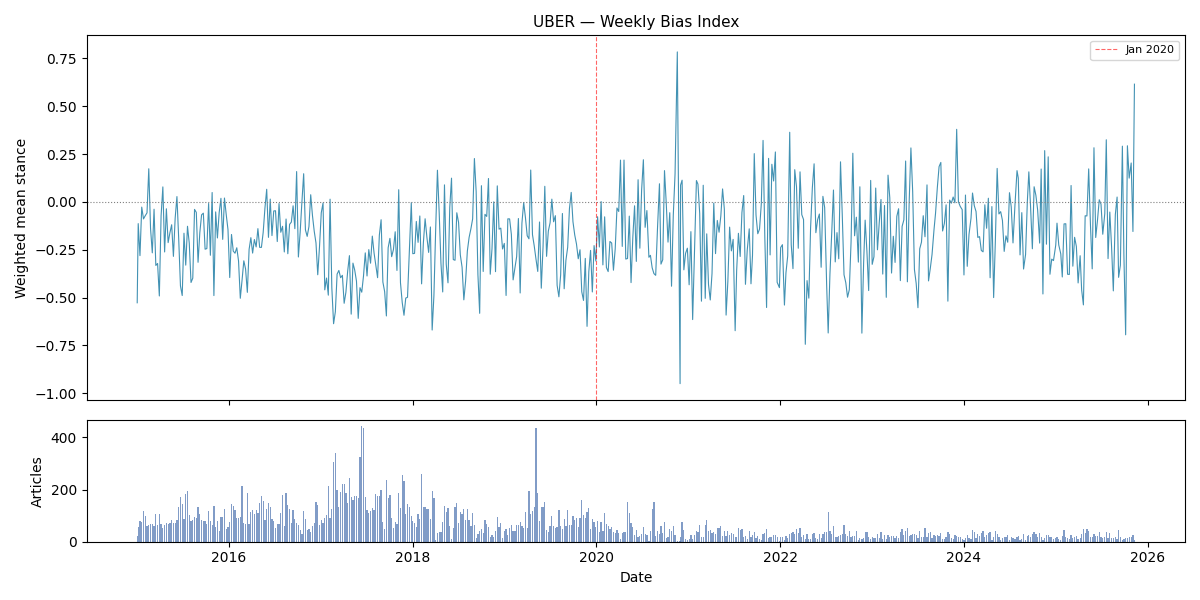}
\caption{Stance signal for Uber over 2015--2025. The upper panel plots the
coverage-weighted mean stance, which drifts around a stable, mildly negative level with
no visible step at January 2020 (dashed line); the lower panel plots weekly headline
volume, whose wide swings are the variation the RQ1 volume control is meant to absorb.}
\label{fig:stance_uber}
\end{figure}

 Estimating \eqref{eq:rq1} on the $27{,}627$ firm-day stance observations, drawn from
$26$ firms across $3{,}982$ trading days, returns the coefficients in
Table~\ref{tab:results}(a). The post-2020 coefficient is $\hat{\beta} = +0.011$ with
an HC1 standard error of $0.030$ ($p = 0.71$), and clustering by day barely disturbs
it ($\text{SE} = 0.030$, $p = 0.72$), so the within-day dependence between firms is
not hiding a result. We cannot reject $H_0^{(1)}$: once each firm is set against its
own history and the day's common movement is held fixed, the tone of coverage does not
step to a new level after the shock. The volume control behaves exactly as its purpose
demands, with a coefficient of $-0.0038$ ($p < 0.001$), so the days on which a firm is
most heavily covered are the days its average tone is pulled back toward neutral, the
mechanical averaging effect the control exists to absorb. The within $R^2$ of $0.002$
is what one expects when a single firm's day-to-day tone is dominated by the noise of
individual headlines. A null here is not a failure to find something; it says the shock
left no common, firm-level mark on tone once the news cycle's own daily swings are
taken out. Figure~\ref{fig:stance_uber} puts the signal on view for one firm across the
whole window: the tone drifts around a stable, mildly negative mean with no step at the
turn of 2020, while the panel beneath it tracks the weekly coverage volume, whose wide
swings are the very variation the volume control absorbs.

\begin{table}[!hbtp]
\caption{Regression results for RQ1 (stance, Eq.~\ref{eq:rq1}) and RQ2 (returns, Eq.~\ref{eq:rq2}). Both fail to reject the null of no post-2020 level shift once common daily (RQ1) and monthly (RQ2) shocks are absorbed.}
\label{tab:results}
\centering
\small
\textbf{(a) RQ1 -- Daily Stance} (firm $+$ day FE, HC1 SE)\\[3pt]
\begin{tabular}{lrrr}
\toprule
Variable & Coef. & SE & $p$ \\
\midrule
Post ($\hat{\beta}$) & $+0.01096$ & $0.02964$ & $0.712$ \\
Article volume       & $-0.00381$ & $0.00049$ & $<0.001$ \\
\midrule
\multicolumn{4}{l}{$n=27{,}627$;\ 26 firms;\ 3,982 days;\ within $R^2=0.0017$} \\
\multicolumn{4}{l}{day-clustered $p_{\text{Post}}=0.717$} \\
\bottomrule
\end{tabular}

\vspace{7pt}
\textbf{(b) RQ2 -- Daily Returns} (firm $+$ month FE, HC1 SE)\\[3pt]
\begin{tabular}{lrrr}
\toprule
Variable & Coef. & SE & $p$ \\
\midrule
Post ($\hat{\beta}$) & $-0.00013$ & $0.00132$ & $0.921$ \\
S\&P 500 return      & $+1.09876$ & $0.01869$ & $<0.001$ \\
VIX                  & $-0.00003$ & $0.00006$ & $0.684$ \\
\midrule
\multicolumn{4}{l}{$n=20{,}537$;\ 26 firms;\ within $R^2=0.316$} \\
\bottomrule
\end{tabular}
\end{table}

\subsection{RQ2: No Detectable Shift in Firm-Level Returns}
\begin{figure}[!t]
\centering
\includegraphics[width=\linewidth]{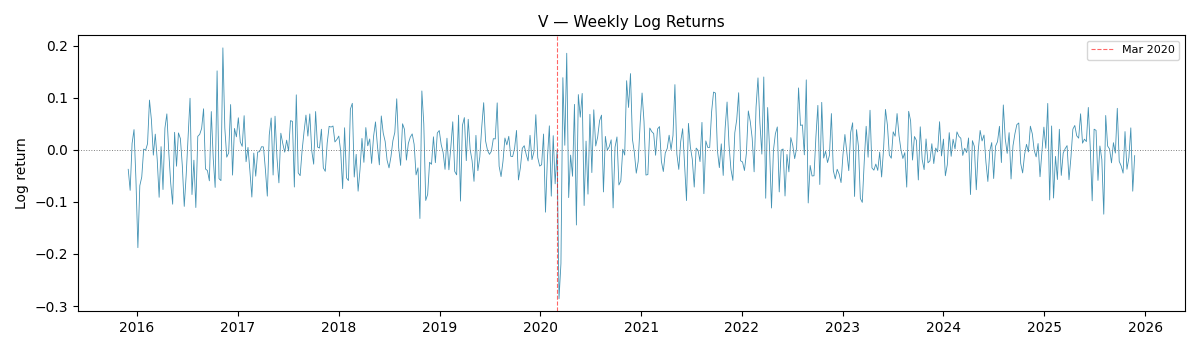}
\caption{Visa weekly log returns, 2015--2025. The March 2020 crash (dashed line) is a
large but short-lived dislocation, after which volatility returns to its earlier range,
consistent with the RQ2 finding of no persistent post-2020 level shift once the market
factor is absorbed.}
\label{fig:returns_v}
\end{figure}

The same estimation on returns, \eqref{eq:rq2} over $20{,}537$ firm-day observations,
tells the matching story (Table~\ref{tab:results}(b)). The post-2020 coefficient is
all but zero, $\hat{\beta} = -0.0001$ ($\text{SE} = 0.0013$, $p = 0.92$): returns show
no level break either. The weekly return series makes the pattern plain
(Figure~\ref{fig:returns_v}): the March 2020 crash is a violent but brief dislocation,
after which volatility settles back toward its earlier range rather than to a new
level. What the regression does register is the market itself. The
S\&P~500 return enters with a coefficient of $1.10$ ($p < 0.001$), the familiar market
beta near one, and together with the monthly effects it accounts for about a third of
the within-firm variation in returns ($R^2 = 0.32$). Volatility adds nothing once the
market return is present ($\text{VIX}$, $p = 0.68$). The reading echoes RQ1: firm
returns move with the market, as they always have, but the shock itself leaves no
separate firm-level step behind. Taken together the two nulls are informative rather
than empty; after common shocks are removed, neither the tone of coverage nor firm
returns jumped in 2020. However, a level test is not the only way the shock could have mattered - it could have changed how tone and returns move \emph{with one another} over
time rather than where each of them sits, and a static regression cannot see that.

\subsection{RQ3: A Link That Lives in Individual Firms}
Run firm by firm on the full sample, the Granger tests are nearly silent, which is
what a regime-dependent relationship would lead us to expect. Only two firms show
stance leading returns, Uber decisively ($p = 0.003$) and Boeing weakly
($p = 0.10$), and the reverse channel, returns leading stance, reaches significance
for Goldman Sachs alone ($p = 0.004$); for every other firm neither direction clears
the five-percent level (Table~\ref{tab:granger}). Taken at face value this reads as a
relationship that is barely there.

\begin{figure}[!t]
\centering
\includegraphics[width=\linewidth]{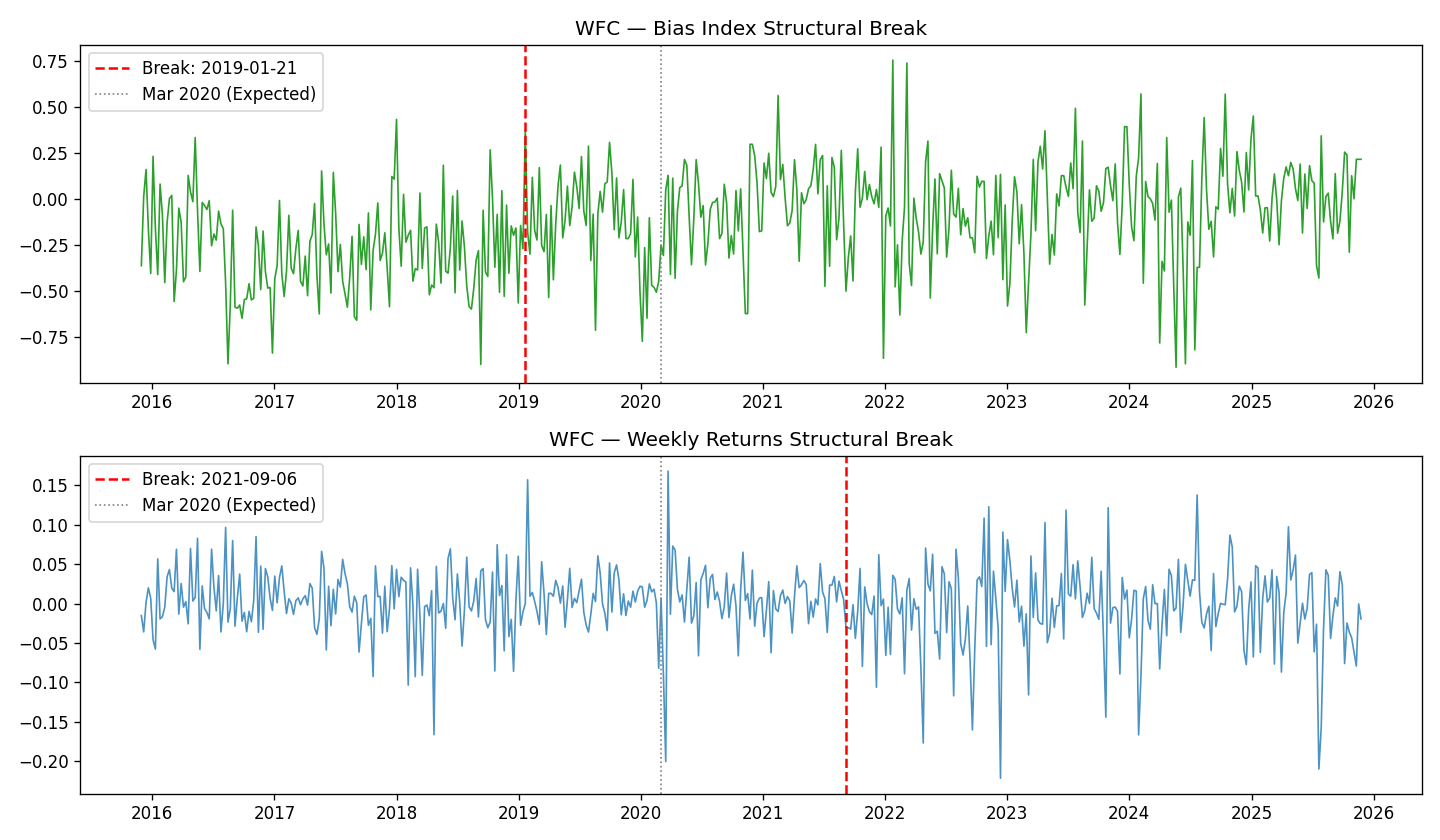}
\caption{Data-driven structural breaks for Wells Fargo. The Bai--Perron break in the
coverage-stance series (top, January 2019) and in weekly returns (bottom, September
2021) both fall well away from the expected March 2020 date (dotted), which is why the
breaks that split the RQ3 subsamples are estimated from each series rather than imposed
on the calendar.}
\label{fig:breaks_wfc}
\end{figure}

Splitting each firm at its own estimated break tells a sharper story. Several firms
that are quiet over the full sample come alive after their break: Uber's
stance-to-returns channel strengthens from $p = 0.48$ before the break to
$p < 0.001$ after it, and Intel turns marginally predictive ($p = 0.074$ post-break).
Feedback from returns to coverage surfaces after the break for AT\&T ($p = 0.015$),
Boeing ($p = 0.010$), and Wells Fargo ($p = 0.030$), while Goldman Sachs and Morgan
Stanley show it only before. This pre/post asymmetry is the firm-level counterpart of
the structural shift the level tests looked for and did not find: the media--market
link is not a fixed feature of the sample but one that switches on after the break for
a subset of firms. The estimated break dates are themselves the argument for reading
them off the data rather than fixing them by hand (Figure~\ref{fig:breaks_wfc}): Wells
Fargo's coverage shifts in early 2019 and its returns only in late 2021, neither near
the pandemic, so a break imposed at March 2020 would have split the firm at the wrong
moment.

Whether those scattered firm-level signals add up to a market-wide channel is the
question the panel VAR settles. With the firms pooled through the Helmert transform,
three lags, and the post-2020 indicator, the VIX, the weekly S\&P~500 return, and
coverage volume all partialled out, neither Granger direction is significant in any
subsample: stance-to-returns and returns-to-stance both sit far from conventional
levels over the full, pre-break, and post-break windows (Table~\ref{tab:panelvar}).
Pooling buys the statistical power the single-firm tests lack, and still nothing
aggregate survives once the market-wide controls are in place. The reading is not that
the link is absent but that it is local: where predictability exists it belongs to
individual firms and their own break dates rather than to the market as a whole. If there were going to be a false positive, it would be expected to appear in the pooled model. However, the lack thereof strengthens the conclusion that the aggregate effect probably doesn't exist.

\begin{table}[!t]
\caption{Firm-level Granger causality on the full sample (HAC $p$-values). Only firms
with a significant direction are shown; the remaining firms clear neither direction at
the five-percent level.}
\label{tab:granger}
\centering
\small
\begin{tabular}{lrrr}
\toprule
Firm & Lag & Stance$\to$Ret & Ret$\to$Stance \\
\midrule
Uber          & 6 & $\mathbf{0.003}$ & $0.375$ \\
Boeing        & 5 & $0.097$          & $0.119$ \\
Goldman Sachs & 3 & $0.432$          & $\mathbf{0.004}$ \\
\bottomrule
\end{tabular}
\end{table}

\begin{table}[!t]
\caption{Panel VAR Granger causality, Helmert-transformed and HAC-robust, with
$\text{Post}_t$, the VIX, the S\&P~500 return, and coverage volume partialled out.
No direction is significant in any subsample.}
\label{tab:panelvar}
\centering
\small
\begin{tabular}{lrrr}
\toprule
Sample & $N$ & Stance$\to$Ret & Ret$\to$Stance \\
\midrule
Full       & $7{,}493$ & $0.565$ & $0.899$ \\
Pre-break  & $4{,}415$ & $0.301$ & $0.861$ \\
Post-break & $3{,}018$ & $0.905$ & $0.937$ \\
\bottomrule
\end{tabular}
\end{table}

\section{Discussion}

\subsection{RQ1: What a Null Stance Shift Means}
The absence of a detectable stance shift implies something non-trivial. The pandemic was inescapable in the news, yet once each firm is set against
its own history and the shocks common to every firm on a day are held aside, the
average direction of its coverage did not shift. The strongly negative volume
coefficient completes the picture: what moved on a firm's heaviest days was how much
was written, not how favorably. For an investor or a regulator worried that the
crisis bent the press systematically for or against large firms, the daily
firm-level evidence offers no support.

The measurement bounds that claim, though it does not undercut it. A signed daily mean is a coarse summary of tone: it collapses opposing headlines from the same day into one number, so a shift in the spread of a firm's coverage rather than in its centre would leave the mean untouched. And a headline, however sharply it compresses a story, is not the full article beneath it, so a stance carried mainly in the body would escape a headline-level measure.

Two extensions follow directly. A distributional summary of daily tone, rather than a signed mean, would keep the within-day spread the mean erases and reveal any change in the shape of coverage the level test cannot. And scoring the full article text for the most heavily covered firms would measure what the headline restriction costs, and recover a stance that lives mainly in the body.

\subsection{RQ2: What a Null Return Shift Means}
For returns the null carries a different message. Once the market factor and the broad
monthly regime are absorbed, the shock left no firm-specific return signature aligned
to the break for the average firm in the panel, which is what efficient pricing of a
market-wide event would predict. 
The pandemic primarily induced a market-wide repricing, while persistent firm-level effects remain largely unexplained after accounting for market controls.

The design is deliberately conservative, and that conservatism is its limitation. By
removing the common component it stays silent on the large aggregate repricing that
plainly happened, and by testing a shift in the mean it cannot see the moves in
volatility and tail risk that were violent in 2020. Loading the broad market onto
monthly effects and the index return also risks absorbing part of a genuine
firm-level response along with the noise. And the panel treats every firm alike,
though the crisis fell on an airline and a grocer in opposite ways.

Each restriction invites a relaxation: modeling conditional volatility directly rather
than testing means alone, defining the firm-specific channel against a richer factor
benchmark than the index and the VIX, and allowing the shock's effect to differ across firms and sectors rather than holding it to one panel-wide coefficient, so that opposing responses no longer average away.

\subsection{RQ3: A Link That Is Firm-Specific, Not Market-Wide}
The dynamic analysis delivers the paper's central interpretive point. Predictability
between stance and returns is sometimes real, but it appears in individual firms and tends to
switch on only after a firm's own structural break; pooled across the panel, with
volatility, the market return, and coverage volume held aside, no market-wide
lead--lag channel survives in either direction. A pooled test is the design most able
to conjure an aggregate effect, and it declines to, so the honest reading is
heterogeneity rather than absence.

The caveats are those of any reduced-form dynamic analysis. Since daily firm-level stance data are sparse, the use of weekly aggregation may mask rapid market responses to news that occur within days or hours. The bivariate design carries only stance and
sector shocks, sit outside it, and the panel controls narrow that gap without closing
it. Additionally, 
Granger causality is predictive, not structural, a rejection says past stance forecasts
returns, not that stance moves prices.

Each caveat points to a natural fix. Pushing to daily or event-time frequency for the densely covered firms would recover any faster channel that weekly aggregation aliases away. Allowing several breaks, or a smoothly switching regime, in place of a single one would fit firms whose dynamics shift more than once. Adding the omitted drivers as exogenous controls would close the gap the bivariate design leaves open.

\section{Conclusion}

This paper set out to examine whether the relationship between financial news and equity
markets changed during the 2020 pandemic. The motivation was twofold. First, while the
efficient market hypothesis suggests that publicly available information should already
be reflected in prices, it remains unclear whether the tone of news coverage carries
predictive information or simply follows movements that have already occurred. Second,
the existing literature largely studies this question through aggregate sentiment
indices, making it difficult to distinguish relationships that are genuinely
market-wide from those that belong only to individual firms. We addressed these
questions by constructing a firm-specific stance measure from $90{,}579$ materially
relevant headlines covering $26$ large United States firms and by combining panel
regressions with firm-level and panel vector autoregressions around data-driven
structural breaks.

The evidence consistently points to the same conclusion. After controlling for
firm-specific heterogeneity and common market shocks, neither media stance nor
firm-level returns exhibit a persistent level shift following the pandemic. Dynamic
relationships between stance and returns do emerge, but they do so only for a limited
number of firms and frequently only after each firm's own estimated structural break.
Once firms are considered jointly and common market influences are removed, no
market-wide lead--lag relationship remains in either direction.

Taken together, these findings suggest that the apparent relationship between financial
news and stock returns is substantially more heterogeneous than much of the existing
literature implies. Studies built on aggregate sentiment can detect market-level
associations, but they cannot determine whether those associations reflect behaviour
shared across firms or the combined effect of a relatively small number of firms with
stronger individual dynamics. By measuring stance at the firm level, estimating
structural breaks from the data rather than imposing them from the calendar, and
examining both firm-level and pooled dynamics, this study shows that the latter
interpretation better explains the evidence. The 2020 shock therefore does not appear
to have fundamentally altered the relationship between financial news and equity
markets, instead it highlights that the relationship is primarily firm-specific,
emerging under particular circumstances rather than as a persistent feature of the
market as a whole.


\bibliographystyle{IEEEtran}
\bibliography{references}

@article{fama1970,
  author  = {Fama, Eugene F.},
  title   = {Efficient Capital Markets: A Review of Theory and Empirical Work},
  journal = {The Journal of Finance},
  volume  = {25},
  number  = {2},
  pages   = {383--417},
  year    = {1970},
  doi     = {10.2307/2325486}
}

@inproceedings{hamborg2021,
  author    = {Hamborg, Felix and Donnay, Karsten},
  title     = {{NewsMTSC}: A Dataset for (Multi-)Target-dependent Sentiment Classification in Political News Articles},
  booktitle = {Proceedings of the 16th Conference of the European Chapter of the Association for Computational Linguistics: Main Volume},
  pages     = {1663--1675},
  year      = {2021},
  doi       = {10.18653/v1/2021.eacl-main.142}
}

@article{costola2023,
  author  = {Costola, Michele and Hinz, Oliver and Nofer, Michael and Pelizzon, Loriana},
  title   = {Machine learning sentiment analysis, {COVID-19} news and stock market reactions},
  journal = {Research in International Business and Finance},
  volume  = {64},
  pages   = {101881},
  year    = {2023},
  doi     = {10.1016/j.ribaf.2023.101881}
}

@article{verma2025,
  author  = {Verma, Rahul and Verma, Priti},
  title   = {Economic News, Social Media Sentiments, and Stock Returns: Which Is a Bigger Driver?},
  journal = {Journal of Risk and Financial Management},
  volume  = {18},
  number  = {1},
  pages   = {16},
  year    = {2025},
  doi     = {10.3390/jrfm18010016}
}

@article{huynh2021,
  author  = {Huynh, Toan Luu Duc and Foglia, Matteo and Nasir, Muhammad Ali and Angelini, Eliana},
  title   = {Feverish sentiment and global equity markets during the {COVID-19} pandemic},
  journal = {Journal of Economic Behavior \& Organization},
  volume  = {188},
  pages   = {1088--1108},
  year    = {2021},
  doi     = {10.1016/j.jebo.2021.06.016}
}

@article{kimhahm2025,
  author  = {Kim-Hahm, Hyunsun and Abou-Zaid, Ahmed S. and Mohd, Abidalrahman},
  title   = {News vs. Social Media: Sentiment Impact on Stock Performance of Big Tech Companies},
  journal = {Journal of Risk and Financial Management},
  volume  = {18},
  number  = {12},
  pages   = {660},
  year    = {2025},
  doi     = {10.3390/jrfm18120660}
}

@article{anastasiou2026,
  author  = {Anastasiou, Dimitris and Ballis, Antonis and Kallandranis, Christos and Vlassas, Ioannis},
  title   = {Positive {COVID-19} related sentiment, economic uncertainty and risk management implications},
  journal = {Journal of Banking Regulation},
  volume  = {27},
  number  = {1},
  pages   = {1--13},
  year    = {2026},
  doi     = {10.1057/s41261-025-00303-z}
}

@article{anastasiou2022,
  author  = {Anastasiou, Dimitris and Ballis, Antonis and Drakos, Konstantinos},
  title   = {Constructing a positive sentiment index for {COVID-19}: Evidence from {G20} stock markets},
  journal = {International Review of Financial Analysis},
  volume  = {81},
  pages   = {102111},
  year    = {2022},
  doi     = {10.1016/j.irfa.2022.102111}
}

@article{eierle2022,
  author  = {Eierle, Brigitte and Klamer, Sebastian and Muck, Matthias},
  title   = {Does it really pay off for investors to consider information from social media?},
  journal = {International Review of Financial Analysis},
  volume  = {81},
  pages   = {102074},
  year    = {2022},
  doi     = {10.1016/j.irfa.2022.102074}
}

@article{sing2023,
  author  = {Sing, Nang Biak and Singh, Rajkumar Giridhari},
  title   = {Investor attention and reaction in {COVID-19} crisis: sentiment analysis in the {Indian} stock market},
  journal = {Managerial Finance},
  volume  = {49},
  number  = {3},
  pages   = {470--491},
  year    = {2023},
  doi     = {10.1108/MF-06-2021-0258}
}

@article{bai2023,
  author  = {Bai, Chenjiang and Duan, Yuejiao and Fan, Xiaoyun and Tang, Shuai},
  title   = {Financial market sentiment and stock return during the {COVID-19} pandemic},
  journal = {Finance Research Letters},
  volume  = {54},
  pages   = {103709},
  year    = {2023},
  doi     = {10.1016/j.frl.2023.103709}
}

@article{dong2022,
  author  = {Dong, Xiuliang and Xu, Shiying and Liu, Jianing and Tsai, Fu-Sheng},
  title   = {Does media sentiment affect stock prices? Evidence from {China's} {STAR} market},
  journal = {Frontiers in Psychology},
  volume  = {13},
  pages   = {1040171},
  year    = {2022},
  doi     = {10.3389/fpsyg.2022.1040171}
}

@article{chen2022,
  author  = {Chen, Chen and Moeini Gharagozloo, M. Mahdi and Darougar, Layla and Shi, Lei},
  title   = {The way digitalization is impacting international financial markets: Stock price synchronicity},
  journal = {International Finance},
  volume  = {25},
  number  = {3},
  pages   = {396--415},
  year    = {2022},
  doi     = {10.1111/infi.12416}
}

@article{ruan2025,
  author  = {Ruan, Linyan and Jiang, Haiwei},
  title   = {Stock Price Prediction Using {FinBERT}-Enhanced Sentiment with {SHAP} Explainability and Differential Privacy},
  journal = {Mathematics},
  volume  = {13},
  number  = {17},
  pages   = {2747},
  year    = {2025},
  doi     = {10.3390/math13172747}
}

@misc{liu2025,
  author       = {Liu, Yiwei and Wang, Junbo and Long, Lei and Li, Xin and Ma, Ruiting and Wu, Yuankai and Chen, Xuebin},
  title        = {A Multi-Level Sentiment Analysis Framework for Financial Texts},
  howpublished = {arXiv preprint arXiv:2504.02429},
  year         = {2025},
  doi          = {10.48550/arXiv.2504.02429}
}

@inproceedings{ronningstad2022,
  author    = {R{\o}nningstad, Egil and Velldal, Erik and {\O}vrelid, Lilja},
  title     = {Entity-Level Sentiment Analysis ({ELSA}): An Exploratory Task Survey},
  booktitle = {Proceedings of the 29th International Conference on Computational Linguistics},
  pages     = {6773--6783},
  address   = {Gyeongju, Republic of Korea},
  year      = {2022}
}

@misc{gyawali2025,
  author       = {Gyawali, Nikesh and Caragea, Doina and Vasenkov, Alex and Caragea, Cornelia},
  title        = {Evaluating Large Language Models for Stance Detection on Financial Targets from {SEC} Filing Reports and Earnings Call Transcripts},
  howpublished = {arXiv preprint arXiv:2510.23464},
  year         = {2025},
  doi          = {10.48550/arXiv.2510.23464}
}

@inproceedings{conforti2022,
  author    = {Conforti, Costanza and Berndt, Jakob and Pilehvar, Mohammad Taher and Giannitsarou, Chryssi and Toxvaerd, Flavio and Collier, Nigel},
  title     = {Incorporating Stock Market Signals for {Twitter} Stance Detection},
  booktitle = {Proceedings of the 60th Annual Meeting of the Association for Computational Linguistics (Volume 1: Long Papers)},
  pages     = {4074--4091},
  address   = {Dublin, Ireland},
  year      = {2022},
  doi       = {10.18653/v1/2022.acl-long.281}
}

@inproceedings{vamvourellis2025,
  author    = {Vamvourellis, Dimitris and Mehta, Dhagash},
  title     = {Reasoning or Overthinking: Evaluating Large Language Models on Financial Sentiment Analysis},
  booktitle = {Proceedings of the 6th ACM International Conference on AI in Finance (ICAIF '25)},
  pages     = {299--307},
  year      = {2025},
  doi       = {10.1145/3768292.3770341}
}

@misc{kirtac2025,
  author       = {Kirtac, Kemal and Germano, Guido},
  title        = {Large language models in finance: what is financial sentiment?},
  howpublished = {arXiv preprint arXiv:2503.03612},
  year         = {2025},
  doi          = {10.48550/arXiv.2503.03612}
}

@misc{kubica2025,
  author       = {Kubica, Dominick and Gordon, Dylan T. and Emura, Nanami and Saini, Derleen and Goldenberg, Charlie},
  title        = {Can {AI} Read Between the Lines? Benchmarking {LLMs} on Financial Nuance},
  howpublished = {arXiv preprint arXiv:2505.16090},
  year         = {2025},
  doi          = {10.48550/arXiv.2505.16090}
}

@inproceedings{tang2023,
  author    = {Tang, Yixuan and Yang, Yi and Huang, Allen and Tam, Andy and Tang, Justin},
  title     = {{FinEntity}: Entity-level Sentiment Classification for Financial Texts},
  booktitle = {Proceedings of the 2023 Conference on Empirical Methods in Natural Language Processing},
  pages     = {15465--15471},
  address   = {Singapore},
  year      = {2023},
  doi       = {10.18653/v1/2023.emnlp-main.956}
}

@article{daudert2022,
  author  = {Daudert, Tobias},
  title   = {A multi-source entity-level sentiment corpus for the financial domain: the {FinLin} corpus},
  journal = {Language Resources and Evaluation},
  volume  = {56},
  pages   = {333--356},
  year    = {2022},
  doi     = {10.1007/s10579-021-09555-3}
}

@misc{gandhi2025,
  author       = {Gandhi, Vishal and Gandhi, Sagar},
  title        = {Prompt Sentiment: The Catalyst for {LLM} Change},
  howpublished = {arXiv preprint arXiv:2503.13510},
  year         = {2025},
  doi          = {10.48550/arXiv.2503.13510}
}

@inproceedings{li2023,
  author    = {Li, Xianzhi and Chan, Samuel and Zhu, Xiaodan and Pei, Yulong and Ma, Zhiqiang and Liu, Xiaomo and Shah, Sameena},
  title     = {Are {ChatGPT} and {GPT-4} General-Purpose Solvers for Financial Text Analytics? A Study on Several Typical Tasks},
  booktitle = {Proceedings of the 2023 Conference on Empirical Methods in Natural Language Processing: Industry Track},
  pages     = {408--422},
  address   = {Singapore},
  year      = {2023},
  doi       = {10.18653/v1/2023.emnlp-industry.39}
}

@article{mahmoudi2022,
  author  = {Mahmoudi, Nader and Docherty, Paul and Melia, Adrian},
  title   = {Firm-level investor sentiment and corporate announcement returns},
  journal = {Journal of Banking \& Finance},
  volume  = {144},
  pages   = {106586},
  year    = {2022}
}

@misc{araci2019,
  author       = {Araci, Dogu},
  title        = {{FinBERT}: Financial Sentiment Analysis with Pre-Trained Language Models},
  howpublished = {arXiv preprint arXiv:1908.10063},
  year         = {2019}
}

@inproceedings{liu2020,
  author    = {Liu, Zhuang and Huang, Degen and Huang, Kaiyu and Li, Zhuang and Zhao, Jun},
  title     = {{FinBERT}: A Pre-trained Financial Language Representation Model for Financial Text Mining},
  booktitle = {Proceedings of the Twenty-Ninth International Joint Conference on Artificial Intelligence (IJCAI-20)},
  pages     = {4513--4519},
  year      = {2020},
  doi       = {10.24963/ijcai.2020/622}
}

@misc{heever2026,
  author       = {van der Heever, Wihan and Ong, Keane and Satapathy, Ranjan and Cambria, Erik},
  title        = {Beyond Correlation: Refutation-Validated Aspect-Based Sentiment Analysis for Explainable Energy Market Returns},
  howpublished = {arXiv preprint arXiv:2603.21473},
  year         = {2026},
  doi          = {10.48550/arXiv.2603.21473}
}

@misc{hu2026,
  author       = {Hu, Xiaoyu and Zhao, Jinman},
  title        = {{Fin-Bias}: Comprehensive Evaluation for {LLM} Decision-Making under Human Bias in the Finance Domain},
  howpublished = {arXiv preprint arXiv:2605.09106},
  year         = {2026},
  doi          = {10.48550/arXiv.2605.09106}
}

@misc{benhenda2026,
  author       = {Benhenda, Mostapha},
  title        = {{Look-Ahead-Bench}: a Standardized Benchmark of Look-ahead Bias in Point-in-Time {LLMs} for Finance},
  howpublished = {arXiv preprint arXiv:2601.13770},
  year         = {2026},
  doi          = {10.48550/arXiv.2601.13770}
}

@misc{eliseev2026,
  author       = {Eliseev, Alexander and Seleznev, Sergei},
  title        = {Fake Date Tests: Can We Trust In-sample Accuracy of {LLMs} in Macroeconomic Forecasting?},
  howpublished = {arXiv preprint arXiv:2601.07992},
  year         = {2026},
  doi          = {10.48550/arXiv.2601.07992}
}

@inproceedings{priya2025,
  author    = {Baghavathi Priya, S. and Kumar, Madhav and Nitheesh Prakash, J. D. and Krithika, N.},
  title     = {Advanced Financial Sentiment Analysis Using {FinBERT} to Explore Sentiment Dynamics},
  booktitle = {Proceedings of the 3rd International Conference on Intelligent Data Communication Technologies and Internet of Things (IDCIoT)},
  pages     = {889--897},
  publisher = {IEEE},
  year      = {2025}
}

@article{pagano2023,
  author  = {Pagano, Marco and Wagner, Christian and Zechner, Josef},
  title   = {Disaster resilience and asset prices},
  journal = {Journal of Financial Economics},
  volume  = {150},
  number  = {2},
  pages   = {103712},
  year    = {2023},
  doi     = {10.1016/j.jfineco.2023.103712}
}

@article{smith2011,
  author  = {Smith, L. Vanessa and Yamagata, Takashi},
  title   = {Firm level return--volatility analysis using dynamic panels},
  journal = {Journal of Empirical Finance},
  volume  = {18},
  number  = {5},
  pages   = {847--867},
  year    = {2011}
}

@article{mamaysky2024,
  author  = {Mamaysky, Harry},
  title   = {News and Markets in the Time of {COVID-19}},
  journal = {Journal of Financial and Quantitative Analysis},
  volume  = {59},
  number  = {8},
  pages   = {3564--3600},
  year    = {2024},
  doi     = {10.1017/S002210902300131X}
}

@article{ballinari2020,
  author  = {Ballinari, Daniele and Behrendt, Simon},
  title   = {Structural breaks in online investor sentiment: A note on the nonstationarity of financial chatter},
  journal = {Finance Research Letters},
  volume  = {35},
  pages   = {101479},
  year    = {2020}
}

@article{moutinho2025,
  author  = {Moutinho, Victor Ferreira and Correia Domingues, Renato Heitor and Fantini, Giulia and Moraes, Michelle},
  title   = {Examining time-varying causality: investor sentiment and asset spreads across {COVID} and {Ukraine} War periods},
  journal = {Applied Economics Letters},
  year    = {2025},
  doi     = {10.1080/13504851.2025.2600676}
}

@article{ditzen2025stata,
  author  = {Ditzen, Jan and Karavias, Yiannis and Westerlund, Joakim},
  title   = {Testing and estimating structural breaks in time series and panel data in {Stata}},
  journal = {The Stata Journal},
  volume  = {25},
  number  = {3},
  pages   = {526--560},
  year    = {2025},
  doi     = {10.1177/1536867X251365449}
}

@article{ditzen2025jae,
  author  = {Ditzen, Jan and Karavias, Yiannis and Westerlund, Joakim},
  title   = {Multiple Structural Breaks in Interactive Effects Panel Data Models},
  journal = {Journal of Applied Econometrics},
  volume  = {40},
  number  = {1},
  pages   = {74--88},
  year    = {2025},
  doi     = {10.1002/jae.3097}
}

@article{karavias2023,
  author  = {Karavias, Yiannis and Narayan, Paresh Kumar and Westerlund, Joakim},
  title   = {Structural Breaks in Interactive Effects Panels and the Stock Market Reaction to {COVID-19}},
  journal = {Journal of Business \& Economic Statistics},
  volume  = {41},
  number  = {3},
  pages   = {653--666},
  year    = {2023},
  doi     = {10.1080/07350015.2022.2053690}
}

@misc{pretis2026,
  author       = {Pretis, Felix and Schwarz, Moritz},
  title        = {Discovering What Mattered: Detecting Unknown Treatment as Breaks in Panel Models},
  howpublished = {SSRN Working Paper 4022745},
  year         = {2026},
  doi          = {10.2139/ssrn.4022745}
}

@article{rossi2019,
  author  = {Rossi, Barbara and Wang, Yiru},
  title   = {Vector autoregressive-based {Granger} causality test in the presence of instabilities},
  journal = {The Stata Journal},
  volume  = {19},
  number  = {4},
  pages   = {883--899},
  year    = {2019},
  doi     = {10.1177/1536867X19893631}
}

@article{lumsdaine2023,
  author  = {Lumsdaine, Robin L. and Okui, Ryo and Wang, Wendun},
  title   = {Estimation of panel group structure models with structural breaks in group memberships and coefficients},
  journal = {Journal of Econometrics},
  volume  = {233},
  number  = {1},
  pages   = {45--65},
  year    = {2023}
}

@article{oliveira2024,
  author  = {de Oliveira, Abdinardo M. B. and Mandal, Anandadeep and Power, Gabriel J.},
  title   = {Impact of {COVID-19} on Stock Indices Volatility: Long-Memory Persistence, Structural Breaks, or Both?},
  journal = {Annals of Data Science},
  volume  = {11},
  number  = {2},
  pages   = {619--646},
  year    = {2024},
  doi     = {10.1007/s40745-022-00446-0}
}

@article{cevik2022,
  author  = {Cevik, Emre and Kirci Altinkeski, Buket and Cevik, Emrah Ismail and Dibooglu, Sel},
  title   = {Investor sentiments and stock markets during the {COVID-19} pandemic},
  journal = {Financial Innovation},
  volume  = {8},
  number  = {1},
  pages   = {69},
  year    = {2022},
  doi     = {10.1186/s40854-022-00375-0}
}

\end{document}